\documentclass[%
 aip,
 amsmath,amssymb,
 reprint,%
]{revtex4-1}

\usepackage{graphicx}
\usepackage{dcolumn}
\usepackage{bm}
\usepackage{xcolor}

\usepackage[utf8]{inputenc}
\usepackage[T1]{fontenc}
\usepackage{mathptmx}

\usepackage{booktabs} 
\usepackage{amsmath}
\allowdisplaybreaks

\begin{document}

\preprint{AIP/123-QED}

\title[]{Unveiling Amplitude Distributions via the Ordinal Language of Random Walks}

\author{A. Mateos Roig}
\affiliation{Department of Statistics and Operations Research, University of Valencia, Doctor Moliner 50, Burjassot, Valencia 46100, Spain}
\affiliation{Instituto de F\'isica Interdisciplinar y Sistemas Complejos IFISC (CSIC-UIB), Campus
UIB, Palma de Mallorca, 07122, Spain}

\author{L. Zunino}%
\affiliation{Centro de Investigaciones \'Opticas (CONICET La Plata-CIC-UNLP), 1897 Gonnet, La Plata, Argentina}
\affiliation{Departamento de Ciencias B\'asicas, Facultad de Ingenier\'ia, Universidad Nacional de La Plata (UNLP), 1900 La Plata, Argentina}

\author{F. Olivares}
\email{olivaresfe@gmail.com}
\affiliation{Instituto de F\'isica Interdisciplinar y Sistemas Complejos IFISC (CSIC-UIB), Campus UIB, Palma de Mallorca, 07122, Spain}

\date{\today}

\begin{abstract}
Ordinal patterns are widely used to characterize temporal organization in time series, yet they are often considered insensitive to the amplitude distribution of the data. In this work, we show that this limitation can be overcome by considering the ordinal structure of integrated time series. We investigate random walks generated from independent non-Gaussian increments and derive analytical expressions for the ordinal pattern probabilities associated with them. We show that for symmetric distributions, the probabilities of some ordinal patterns are fully determined by symmetry arguments, while those for the remaining patterns depend explicitly on the shape of the increments distribution. Numerical simulations based on $q$-Gaussian increments validate the theoretical predictions. We further show that the construction of the random walk itself plays a fundamental role in the accurate characterization of non-Gaussian fluctuations, as different centering procedures may significantly affect the resulting ordinal statistics. Finally, we validate the proposed framework using financial time series, showing that the ordinal distributions of integrated logarithmic returns capture non-Gaussian features consistent with a cubic law. 
\end{abstract}

\maketitle

\begin{quotation}
Many natural and human-made systems exhibit non-Gaussian fluctuations, yet identifying their statistical properties remains challenging because existing methods often require large datasets and are sensitive to noise and outliers. In this work, we show that integrating a signal before applying ordinal pattern analysis transforms its amplitude statistics into geometric constraints on trajectories, allowing the underlying distribution to be inferred from ordinal probabilities. This simple idea reveals information that is normally hidden by ordinal methods, enabling the robust identification of heavy tails and asymmetries. The proposed framework is validated analytically, numerically, and through empirical analyses of financial market data. These results broaden the scope of ordinal analysis and provide a general framework for studying non-Gaussian fluctuations in a wide range of complex systems.

\end{quotation}

\section{Introduction}
Many natural and man-made systems show fluctuations that differ significantly from Gaussian statistics~\cite{clauset2009power,metzler2000random,tsallis2009introduction}. Heavy-tailed distributions, intermittency, and extreme events have been observed in various contexts, including turbulence~\cite{castaing1990velocity,beck2001dynamical}, physiological series~\cite{peng1993long}, flights delays~\cite{mitsokapas2021statistical,olivares2025quantifying} and financial markets~\cite{mandelbrot1963new,gopikrishnan1998inverse,nayak2021computational,gabaix2003theory}. These findings suggest that large deviations happen more often than the central limit theorem predicts. 
Characterizing non-Gaussian fluctuations is a key problem when complex systems are analyzed. Traditional methods usually depend on moments, cumulants, tail exponents, or goodness-of-fit tests to measure deviations from Gaussian distributions. However, these measures often need large datasets and may become unreliable with heavy tails, finite-size effects, or extreme events~\cite{clauset2009power,metzler2000random,tsallis2009introduction}. As a result, developing new methods that can extract information from non-Gaussian signals is an ongoing area of research.

The dynamical behavior of complex systems has been extensively studied through time series analysis and information-theoretic metrics~\cite{tang2015complexity}. Among the available methods, symbolic representations based on ordinal patterns have gained popularity since the early work by Bandt and Pompe~\cite{bandt2002permutation}. By using relative orderings instead of numerical values, ordinal methods offer a strong and efficient framework for time series analysis. Over the last two decades, they have been used effectively to characterize the underlying dynamical nature in data~\cite{rosso2007distinguishing, zanin2021ordinal}, detect temporal correlations~\cite{olivares2016quantifying,olivares2020multiscale}, identify dynamic transitions~\cite{zunino2022permutation}, characterize relevant temporal scales~\cite{zunino2010permutation,soriano2011distinguishing,zunino2012distinguishing,olivares2020multiscale,soriano2021time,zunino2024identifying} and measure temporal irreversibility~\cite{zanin2018assessing,zunino2022permutation,zanin2024manipulating}, in a wide range of natural systems.

One of the most frequently discussed characteristics of ordinal patterns is that they discard amplitude information by construction~\cite{bandt2002permutation}. While this property is advantageous in terms of robustness, it may also imply the loss of relevant information in systems where the magnitude of fluctuations carries dynamical significance~\cite{cuesta2019permutation}. Several extensions of the original framework have been proposed to partially incorporate amplitude effects, including weighted, fine-grained, and adaptive ordinal approaches~\cite{xiao2009fine,fadlallah2013weighted,azami2016amplitude,zanin2023continuous,tyloo2025including}. However, these methodologies mainly focus on improving classification performance or enhancing sensitivity to abrupt changes and spikes~\cite{cuesta2019permutation}. Consequently, the fundamental relationship between ordinal statistics and the underlying amplitude distribution remains poorly understood.

While ordinal patterns extracted from independent and identically distributed (i.i.d.) processes are known to be equiprobable regardless of the underlying distribution, recent studies have shown that this property does not hold for integrated processes or random walks~\cite{deford2017random,olivares2025quantifying}. In particular, the temporal accumulation of independent increments introduces geometric constraints on the resulting trajectories, leading to non-uniform ordinal pattern distributions whose probabilities depend on the statistics of the increments. This observation suggests that, although ordinal patterns computed directly from a signal may be insensitive to its amplitude distribution, the ordinal structure of the corresponding integrated process can retain information about distributional properties that would otherwise remain hidden.

In this work, we investigate how non-Gaussian properties of a probability distribution are encoded in the ordinal structure of the corresponding random walk. We derive analytical expressions for the ordinal probabilities of several random walks and identify a subset of patterns whose occurrence frequencies are completely determined by symmetry, independently of the underlying distribution. We further show that the probability of the remaining patterns become sensitive to higher-order distributional properties, including heavy tails, compact support and asymmetry. Through numerical simulations based on $q$-Gaussian increments and empirical analyses of financial time series, we demonstrate that ordinal pattern distributions from integrated processes provide information about the amplitude distribution of the associated increments. 

  
\section{Ordinal pattern probability distribution}

Ordinal patterns are characterized by two integer parameters: the embedding dimension (or pattern length) $D \geq 2$, and the time delay $\tau \geq 1$, which determines the temporal spacing between consecutive elements of the pattern. Given a one-dimensional time series $X(t) = \{x_t \, ; \, t = 1, \dots, M\}$, the method consists of constructing $D$-dimensional delay vectors of the form
\begin{equation}
\mathbf{v}_t = (x_t, x_{t+\tau}, \dots, x_{t+(D-1)\tau}),
\end{equation}
which may contain consecutive observations when $\tau = 1$, or non-consecutive values when $\tau > 1$. Each vector thus captures the local dynamical structure of the time series across a temporal window of size $(D-1)\tau$. For each vector $\mathbf{v}_t$, the ordinal pattern is obtained by assigning to each temporal position the rank occupied by its value within the vector. The smallest value receives rank 0, the second smallest rank 1, and so on up to rank $D-1$ for the largest value. The resulting sequence of ranks defines a permutation $\pi_i$ of the symbols $\{0,1,\dots,D-1\}$, referred to as the ordinal pattern at time $t$. In this way, the original time series is mapped into a symbolic sequence of permutations encoding the relative ordering of values rather than their absolute magnitudes. When the underlying process is continuous, equal values occur infrequently~\cite{cuesta2018patterns}. However, in empirical datasets with finite resolution or discrete measurements, ties may appear frequently. In such cases, a common strategy consists of breaking ties by adding a small amount of noise to the data prior to the ordinal transformation~\cite{bandt2002permutation}.

The ordinal pattern probability distribution is obtained by counting the number of occurrences of each permutation $\pi_i$ in the symbolized time series and normalizing by the total number of embedded vectors, $M-(D-1)\tau$. Formally,
\begin{equation}
\label{Eq:PDF}
p_i = \frac{\#(\pi_i)}{M-(D-1)\tau}, \qquad i = 1,2,\dots,D!,
\end{equation}
where $\#(\pi_i)$ denotes the number of times the permutation $\pi_i$ appears in the series.

To illustrate this procedure, consider a simple time series of length $M=10$, $X = \{2,6,9,13,8,15,1,18,3,21\}$, with embedding dimension $D=3$ and delay $\tau=1$. In this case, $M-(D-1)\tau = 8$ ordinal patterns are generated. The first two triplets, $(2,6,9)$ and $(6,9,13)$, are mapped to the permutation $(012)$, since their elements are already in ascending order. The triplets $(9,13,8)$ and $(8,15,1)$ correspond to $(120)$, as they satisfy $x_{t+2} < x_t < x_{t+1}$. Similarly, $(13,8,15)$, $(15,1,18)$, and $(18,3,21)$ are mapped to $(102)$ because $x_{t+1} < x_t < x_{t+2}$. Finally, the triplet $(1,18,3)$ yields the permutation $(021)$. The resulting ordinal distribution is therefore
\[
p(012)=\frac{1}{4}, \quad
p(120)=\frac{1}{4}, \quad
p(102)=\frac{3}{8}, \quad
p(021)=\frac{1}{8},
\]
while $p(201)=p(210)=0$.

Since a $D$-dimensional vector admits $D!$ possible permutations, reliable estimation of the ordinal distribution requires the condition $M \gg D!$ to be satisfied~\cite{bandt2002permutation}. Consequently, the maximum admissible pattern length is constrained by the size of the available time series. In their seminal work~\cite{bandt2002permutation}, Bandt and Pompe originally proposed fixing the delay to $\tau = 1$. However, subsequent studies have shown that considering lagged embeddings, i.e., $\tau \geq 2$, can provide additional insight into the underlying dynamics of certain systems. Physically, increasing $\tau$ corresponds to probing the system at multiples of the sampling interval $\delta t$, thereby allowing the exploration of dynamics at different temporal scales~\cite{zunino2010permutation,soriano2011distinguishing,zunino2012distinguishing,olivares2020multiscale,soriano2021time,zunino2024identifying}.

\section{Permutation Jensen-Shannon distance}

The Permutation Jensen–Shannon Distance (PJSD) was recently introduced as a versatile metric to quantify the degree of similarity between the symbolic ordinal statistics of two time series~\cite{zunino2022permutation}. It is defined in terms of the Jensen–Shannon divergence~\cite{lin1991divergence} between two ordinal pattern probability distributions 
$P = \{p_1, \dots, p_N\}$ and $Q = \{q_1, \dots, q_N\}$ associated with the corresponding time series, namely
\begin{equation}
\label{eq1}
D_{\mathrm{JS}}(P,Q) 
= S\!\left(\frac{P+Q}{2}\right) 
- \frac{S(P)}{2} 
- \frac{S(Q)}{2},
\end{equation}
where 
\begin{equation}
S(P) = -\sum_{i=1}^{N} p_i \ln p_i
\end{equation}
denotes the Shannon entropy. The PJSD is obtained by taking the square root of Eq.~(\ref{eq1}), and its normalized form reads
\begin{equation}
\label{eq2}
\mathrm{PJSD}(P,Q) 
= \sqrt{\frac{D_{\mathrm{JS}}(P,Q)}{\ln 2}}.
\end{equation}

As a symmetric and bounded measure of distinguishability between two probability distributions, the PJSD allows a quantitative comparison of the ordinal composition of different time series. Larger values indicate greater dissimilarity between the symbolic representations of the signals, whereas smaller values reflect higher similarity. 

Intuitively, time series generated by the same underlying dynamics are expected to yield small PJSD values (close to, but not exactly, zero due to finite-size effects). Conversely, signals arising from different dynamical mechanisms should produce significantly larger values. Indeed, in the former case, the PJSD has been shown to asymptotically converge to zero as the series length increases~\cite{zunino2022permutation}.
\section{Analytical and Numerical Analysis}

Let $\{x_t\}$ be a sequence of independent fluctuations drawn from a probability density function $f(x)$. In the absence of temporal correlations, the variables are independent and identically distributed (i.i.d.), such that the time series does not contain any temporal organization beyond the statistical properties encoded in the distribution itself. If ordinal patterns are computed directly from the fluctuations $\{x_t\}$, the resulting ordinal probabilities are completely insensitive to the underlying amplitude distribution. Indeed, for any continuous i.i.d. process, all ordinal patterns appear with equal probability, i.e., $p_i=1/D!$. This occurs because, in the absence of temporal correlations, all possible orderings among consecutive independent samples are equally likely. Consequently, the direct ordinal analysis of $\{x_t\}$ remains blind to the shape of the probability distribution itself~\cite{bandt2002permutation}.

To extract information associated with the distribution of fluctuations, we instead analyze the cumulative process (or random walk) generated by the series,
\begin{equation}\label{walk}
    y_t=\sum_{i=1}^{t} (x_i - \langle x \rangle).
\end{equation}
In this representation, the ordinal patterns are not determined solely by the instantaneous ordering of independent samples, but by the geometry of partial sums of the fluctuations. As a consequence, the probabilities of ordinal patterns become sensitive to the statistical properties of $f(x)$, even when the original fluctuations are temporally uncorrelated.

Intuitively, the integration introduces statistical dependencies between consecutive points of the walk,
\begin{equation}
    y_{t+1}=y_t+x_{t+1}-\langle x \rangle,
\end{equation}
such that the relative ordering among neighboring values depends on the accumulated contribution of the fluctuations. The probability of observing a pattern \(\pi\) of length $D$ can be expressed as an integral over \(D-1\) variables, since there are \(D-1\) successive increments ("jumps") in the pattern $\pi$. It reads
\begin{equation}
P(\pi_i) = \int_{\mathcal{S}_{\pi_i}} \left( \prod_{k=1}^{D-1} f(x_k) \right)\, dx_1 \cdots dx_{D-1}.
\end{equation}
Here, \(\mathcal{S}_{\pi_i} \subset \mathbb{R}^{D-1}\) denotes the region of values for which the cumulative jumps reconstruct the specific ordering \(\pi_i\). Note that although we integrate over \(D-1\) variables, these correspond to jumps generated by \(D\) time series values. Each jump is treated as an independent realization from the distribution \(f\). This integral formulation highlights how different patterns may have distinct probabilities, depending on the properties of \(f(x)\), such as symmetry, skewness, or tail behavior.

\subsection{Symmetric Distributions}

Let ${x_t}$ be a sequence of i.i.d random variables drawn from a continuous probability density function $f(x)$, assumed to be symmetric about the origin, i.e., $f(x)=f(-x)$, which implies $\mathbb{E}[x_t]=0$.

\subsubsection{Case $D=3$}
For an embedding dimension $D=3$ and $\tau=1$, ordinal patterns are defined from triplets of consecutive observations. Therefore, in the case of the integrated process, three consecutive points of the walk can be expressed as

\begin{equation}
    y_0=0,\qquad y_1=x_1,\qquad y_2=x_1+x_2,
\end{equation}
where, without loss of generality, we have set the initial position to zero. All six ordinal patterns are determined by the relative ordering of $\{y_0,y_1,y_2\}$. Let's consider the ascending pattern
\begin{equation}
    y_0<y_1<y_2 .
\end{equation}
Using the definitions above, this condition becomes
\begin{equation}
    0<x_1<x_1+x_2 .
\end{equation}
This implies
\begin{equation}
    x_1>0,
    \qquad
    x_2>0.
\end{equation}
Hence, the probability of this ordinal pattern is
\begin{equation}
    P(012)
    =
    \int_{0}^{\infty}\int_{0}^{\infty}
    f(x_1)f(x_2)\,dx_2\,dx_1 .
\end{equation}
Since the distribution is symmetric $\int_{0}^{\infty} f(x)\,dx=\frac{1}{2}$, we obtain
\begin{equation}
    P(012)=\frac{1}{4}.
\end{equation}
By symmetry $P(210)=\frac{1}{4}$. The remaining four patterns correspond to cases in which the walk changes direction. For instance, the pattern $(021)$ is defined by the condition
\begin{equation}
    y_0<y_2<y_1,
\end{equation}
that is equivalent to
\begin{equation}
    0<x_1+x_2<x_1.
\end{equation}
This implies
\begin{equation}
    x_1>0,
    \qquad
    -x_1<x_2<0.
\end{equation}
Therefore,
\begin{equation}
    P(021)
    =
    \int_{0}^{\infty}
    \int_{-x_1}^{0}
    f(x_1)f(x_2)\,dx_2\,dx_1 .
\end{equation}
Using the symmetry of $f(x)$, this can be written as
\begin{equation}
    P(021)
    =
    \int_{0}^{\infty}
    f(x_1)
    \int_{0}^{x_1}
    f(u)\,du\,dx_1 .
\end{equation}
Introducing the cumulative distribution on the positive half-axis,
\begin{equation}
    F_+(x)=\int_{0}^{x}f(u)\,du,
\end{equation}
we obtain
\begin{equation}
    P(021)
    =
    \int_{0}^{\infty} f(x)F_+(x)\,dx .
\end{equation}
Since
\begin{equation}
    \frac{d}{dx}F_+^2(x)=2f(x)F_+(x),
\end{equation}
it follows that
\begin{equation}
    P(021)
    =
    \frac{1}{2}
    \left[
    F_+^2(x)
    \right]_{0}^{\infty}
    =
    \frac{1}{2}
    \left(\frac{1}{2}\right)^2
    =
    \frac{1}{8}.
\end{equation}
By symmetry, the other direction-changing patterns have the same probability,
\begin{equation}
    P(021)=P(102)=P(120)=P(201)=\frac{1}{8}.
\end{equation}
Consequently, for any continuous symmetric distribution (Gaussian or non-Gaussian) the ordinal pattern probabilities of the associated random walk for $D=3$ are
\begin{equation}\label{probD3A}
    P(012)=P(210)=\frac{1}{4},
\end{equation}
and
\begin{equation}\label{probD3B}
    P(021)=P(102)=P(120)=P(201)=\frac{1}{8}.
\end{equation}

\subsubsection{Case $D=4$}
Without loss of generality, we set
\begin{equation}\label{process}
y_0 = 0, \quad y_1 = x_1, \quad y_2 = x_1 + x_2, \quad y_3 = x_1 + x_2 + x_3.
\end{equation}
The ordinal pattern $(0132)$ corresponds to the ordering
\begin{equation}
y_0 < y_1 < y_3 < y_2,
\end{equation}
which translates into the conditions
\begin{equation}
x_1 > 0, \quad x_3 < 0, \quad x_2 > -x_3.
\end{equation}
Assuming that the increments are drawn from a symmetric density $f(x)=f(-x)$, the probability of this pattern can be written as
\begin{equation}
P(0132) = \int_0^\infty f(x_1)\,dx_1 \int_{-\infty}^0 f(x_3) \int_{-x_3}^{\infty} f(x_2)\,dx_2\,dx_3.
\end{equation}
By symmetry, $\int_0^\infty f(x)\,dx = 1/2$, and performing the change of variables $u=-x_3$, one obtains
\begin{equation}
P(0132) = \frac{1}{2} \int_0^\infty f(u) \int_u^\infty f(x_2)\,dx_2\,du.
\end{equation}
The double integral represents the probability that two positive variables, $u$ and $x_2$, satisfy $0< u < x_2$. Since both variables are independent and identically distributed, the joint probability over the positive domain factorizes as
\begin{equation}
P(u>0,x_2>0)=\left(\int_0^\infty f(x)\,dx\right)^2.
\end{equation}
Within this domain, the regions defined by $u<x_2$ and $x_2<u$ are symmetric and therefore equally likely. As the probability of the boundary $u=x_2$ is zero for continuous distributions, each region contributes exactly half of the total probability. Hence,
\begin{equation}
P(0<u<x_2)=\frac{1}{2}\left(\int_0^\infty f(x)\,dx\right)^2.
\end{equation}
By symmetry within the positive domain, this probability is $1/2$, yielding
\begin{equation}
\int_0^\infty f(u) \int_u^\infty f(x_2)\,dx_2\,du = \frac{1}{8}.
\end{equation}
Therefore,
\begin{equation}
P(0132) = \frac{1}{16}.
\end{equation}
This result follows purely from symmetry, as the integration domain involves only adjacent increments and does not require combining multiple variables in the limits. By the same reasoning, all ordinal patterns whose defining inequalities involve only adjacent increments can be evaluated analytically. Further details on these patterns and the explicit integral expressions are provided in Appendix~\ref{Apendice}. Moreover, exploiting the symmetry of the underlying distribution, patterns can be grouped into equivalence classes with identical probabilities. Accordingly, one obtains:
\begin{equation}\label{patrones1}
\begin{aligned}
&P(0123) = P(3210) = \frac{1}{8}, \,\,\,\,\,\,
P(0132) = P(2310) = \frac{1}{16}, \\
&P(0213) = P(3120) = \frac{1}{24}, \,\,\,
P(0312) = P(2130) = \frac{1}{48}, \\
&P(1023) = P(3201) = \frac{1}{16}, \,\,\,
P(1203) = P(3021) = \frac{1}{48}.
\end{aligned}
\end{equation}
Note that, for this subset of patterns, symmetry alone fully determines the probabilities, independently of the specific form of the underlying distribution.

While symmetry arguments determine a subset of pattern probabilities independently of the underlying distribution, the remaining cases depend explicitly on the form of the density function. In general, closed form expressions are not available for arbitrary distributions. However, for specific choices such as the uniform or Laplace distributions, the integrals can still be evaluated analytically due to the simple structure of the density function. In the particular case of uniformly distributed increments, $x_i \sim \mathcal{U}(-a,a)$, the probability reduces to the normalized volume of a region within a bounded domain, since the density function is constant. Therefore, while symmetry alone is not sufficient to determine the result, the integral can still be computed exactly by geometric arguments. On the other hand, the exponential form of the Laplace distribution allows for tractable integrations, in contrast to more complex cases such as heavy-tailed power-law distributions, where numerical evaluation is typically required as we will see below.

Let us again consider four consecutive points of the integrated process shown in Eq.~(\ref{process}). For the ordinal pattern $(0231)$, the ordering condition from the smallest to the largest value is
\begin{equation}
y_t<y_{t+3}<y_{t+1}<y_{t+2}.
\end{equation}
Substituting the integrated variables gives
\begin{equation}
0 < x_1+x_2+x_3 < x_1 < x_1+x_2 .
\end{equation}
From these inequalities we obtain
\begin{equation} \label{inq_0231}
\begin{aligned}
x_1 &> 0, \\
x_2 &> 0, \\
x_3 &< -x_2, \\
x_3 &> -(x_2 + x_1).
\end{aligned}
\end{equation}
Accordingly, the probability can be written as
\begin{equation}\label{eq_int_0231}
P(0231)
=
\int_0^\infty f(x_1)
\int_0^\infty f(x_2)
\int_{-(x_1 + x_2)}^{-x_2} f(x_3)
\,dx_1\,dx_2\,dx_3 .
\end{equation}
At this stage, the integration limits involve combinations of variables, which prevents a decomposition of the domain into symmetric subregions. As a consequence, symmetry arguments alone are no longer sufficient to determine the probability. In contrast to the previous cases, the value of $P(0231)$ depends explicitly on the functional form of the density $f(x)$ and, in general, must be evaluated numerically. The same situation applies to the following ordinal patterns, whose probabilities cannot be determined from symmetry arguments alone and therefore require explicit numerical evaluation:
\begin{equation}
\begin{aligned}
&(0231),\ (0321),\ (1032),\ (1230),\ (1302),\ (1320),\\
&(2013),\ (2031),\ (2103),\ (2301),\ (3012),\ (3102).
\end{aligned}
\end{equation}
The complete set of explicit triple integral representations is provided in Appendix~\ref{Apendice}. 

A dependence of these same ordinal patterns on the underlying distribution has also been reported in synthetic time series generated from the cumulative sums of stable-distributed increments~\cite{olivares2025quantifying}. These findings further support the idea that such patterns encode statistical information about the amplitude distribution of the increments through the geometry of the associated integrated process.

\subsubsection{Numerical analysis with $q$-Gaussian distributed increments}

In order to validate the analytical results derived above, we perform a numerical analysis based on synthetic time series whose increments are drawn from $q$-Gaussian distributions~\cite{tsallis1988possible}. The $q$-Gaussian distribution constitutes a generalization of the standard Gaussian distribution within the framework of nonextensive statistical mechanics~\cite{tsallis2009introduction}. It is defined by the probability density function
\begin{equation}
p_q(x) = A_q \left[1 - (1 - q) B_q x^2 \right]_{+}^{\frac{1}{1-q}},
\end{equation}
where $[z]_+ = \max(z,0)$ ensures that the support of the distribution is restricted to real-valued arguments. The coefficients $A_q$ and $B_q$ are normalization and scale parameters, respectively, that depend on the value of $q$. This distribution allows us to systematically explore the impact of deviations from Gaussianity, ranging from compact-support distributions ($q<1$) to heavy-tailed regimes with power-law decay ($1<q<3$). In the limit $q \to 1$, the $q$-Gaussian recovers the standard Gaussian distribution. 

The sampling is performed using a generalized version of the Box--Muller transform~\cite{thistleton2007generalized}, which allows for an efficient and accurate generation of $q$-Gaussian random variables across both compact-support and heavy-tailed regimes, with zero mean and unit variance (for $q>5/3$, ensuring finite second moment). For each value of $q \in \{-1,-0.5,0,0.5,1,1.5,2\}$, we generate realizations of length $N=10^5$ data points of the increment process and construct the associated integrated signal given by Eq.~(\ref{walk}). Ordinal patterns of embedding dimension $D=4$ and delay $\tau=1$ are then computed from the integrated series. The corresponding probabilities are estimated from relative frequencies and averaged over $20$ independent realizations in order to reduce statistical fluctuations. 

\begin{figure}[t!]
    \centering
    \includegraphics[width=0.5\textwidth]{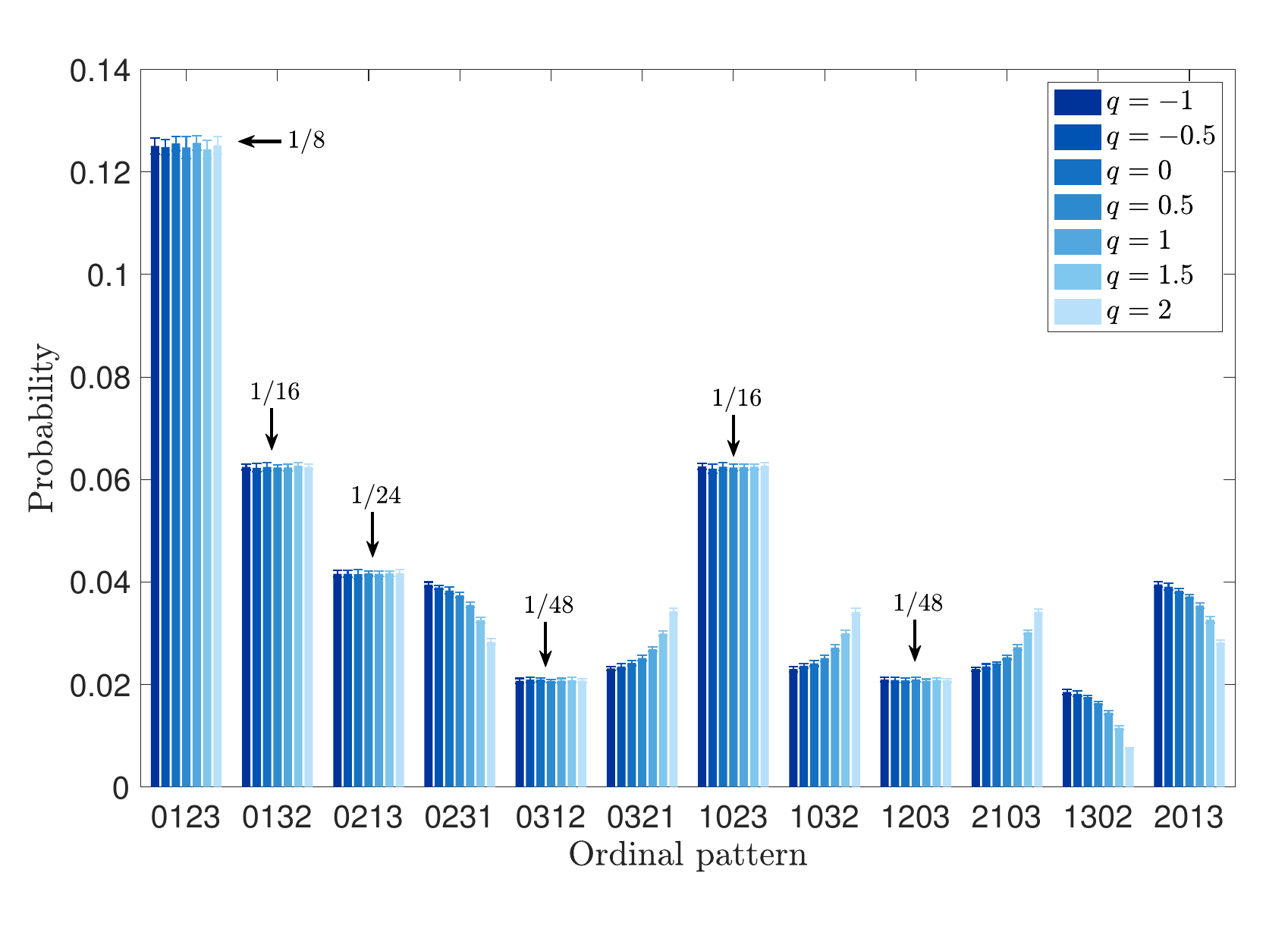}
    \caption{Ordinal pattern probabilities for the integrated process generated from independent $q$-Gaussian distributed increments, as a function of the parameter $q \in [-1,2]$ with a step of $0.5$. Mean and standard deviation (as errorbars) over 20 independent realizations are shown. The theoretical values of the ordinal probabilities given in Eq.~(\ref{patrones1}) are indicated.}
    \label{fig:pdfd4}
\end{figure}

These results are shown in Fig.~\ref{fig:pdfd4}. Due to time-reversal symmetry considerations, only half of the ordinal probability distribution is shown. In particular, inverse ordinal patterns possess identical probabilities, implying that the remaining part of the distribution can be obtained as a mirror image of the displayed one. We observe that the ordinal patterns whose probabilities can be obtained analytically from symmetry arguments alone remain invariant across all explored values of $q$. In other words, these symmetry-constrained patterns are unaffected by the specific shape of the underlying increment distribution, including both compact-support and heavy-tailed regimes. Furthermore, the numerical estimations are found to be in excellent agreement with the theoretical values derived in the previous section, thereby validating the analytical predictions (see Eq.~(\ref{patrones1})).

\begin{figure}[t!]
    \centering
    \includegraphics[width=0.5\textwidth]{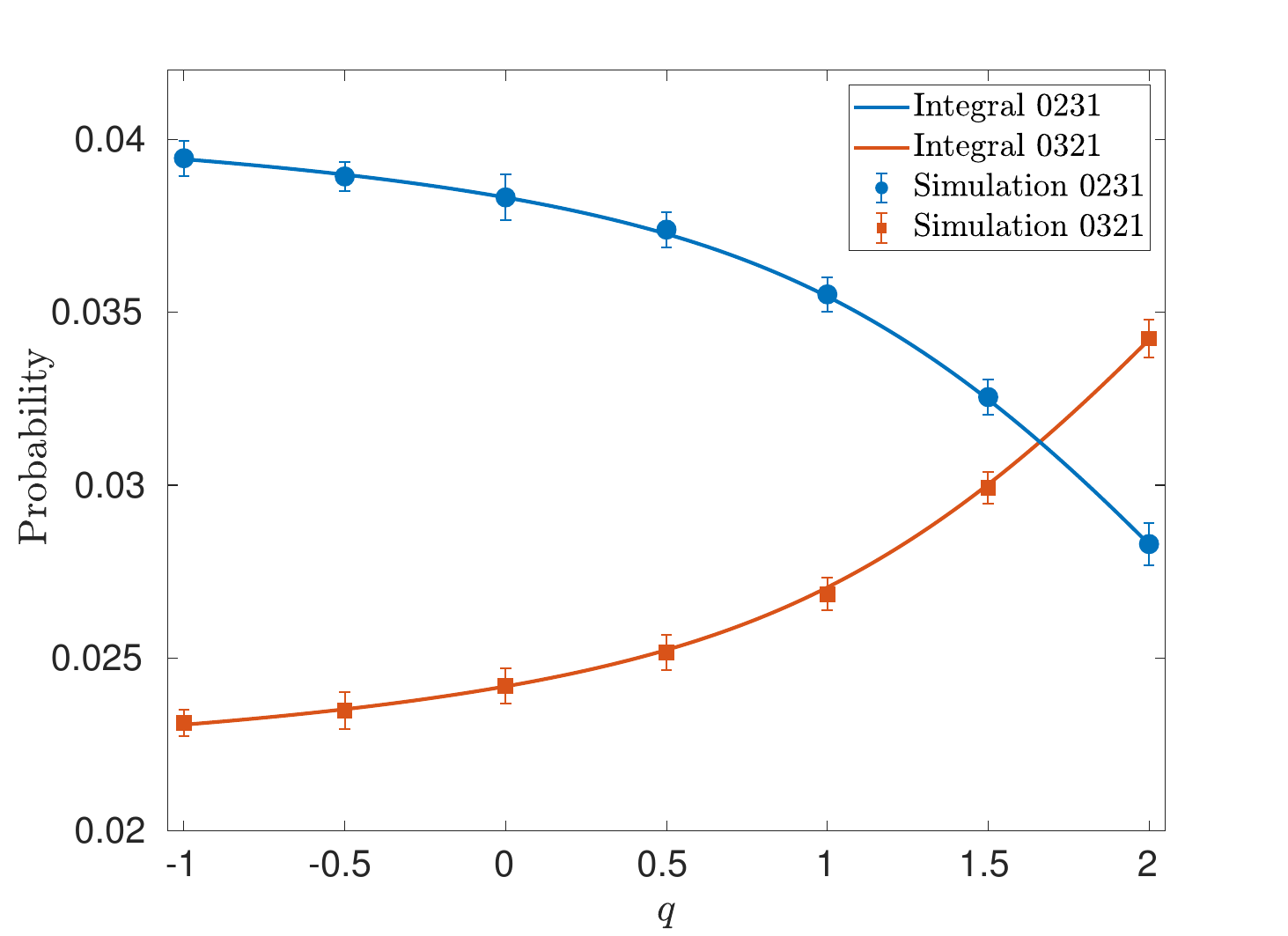}
    \caption{Ordinal pattern probabilities for the integrated process generated from independent $q$-Gaussian distributed increments, as a function of the parameter $q$. Blue dots and red squares represent the probabilities estimated from simulated time series for the patterns $(0231)$ and $(0321)$, respectively. Mean and standard deviation (as errorbars) over 20 independent realizations are shown. The solid blue line corresponds to the numerical integration of Eq.~(\ref{eq_int_0231}), while the solid red line is obtained from the numerical evaluation of $P(0321)
=
\int_{0}^{\infty}
\int_{-x_1}^{0}
\int_{-(x_1+x_2)}^{0} f(x_1)f(x_2)f(x_3)
\,dx_3\,dx_2\,dx_1$. Numerical evaluation of the triple integrals was performed using MATLAB's \texttt{integral3} function, which employs an adaptive quadrature algorithm, with absolute and relative error tolerances set to $\mathrm{AbsTol}=\mathrm{RelTol}=10^{-8}$.}
    \label{fig:sim_teo}
\end{figure}

\begin{figure}[t!]
    \centering
    \includegraphics[width=0.4\textwidth]{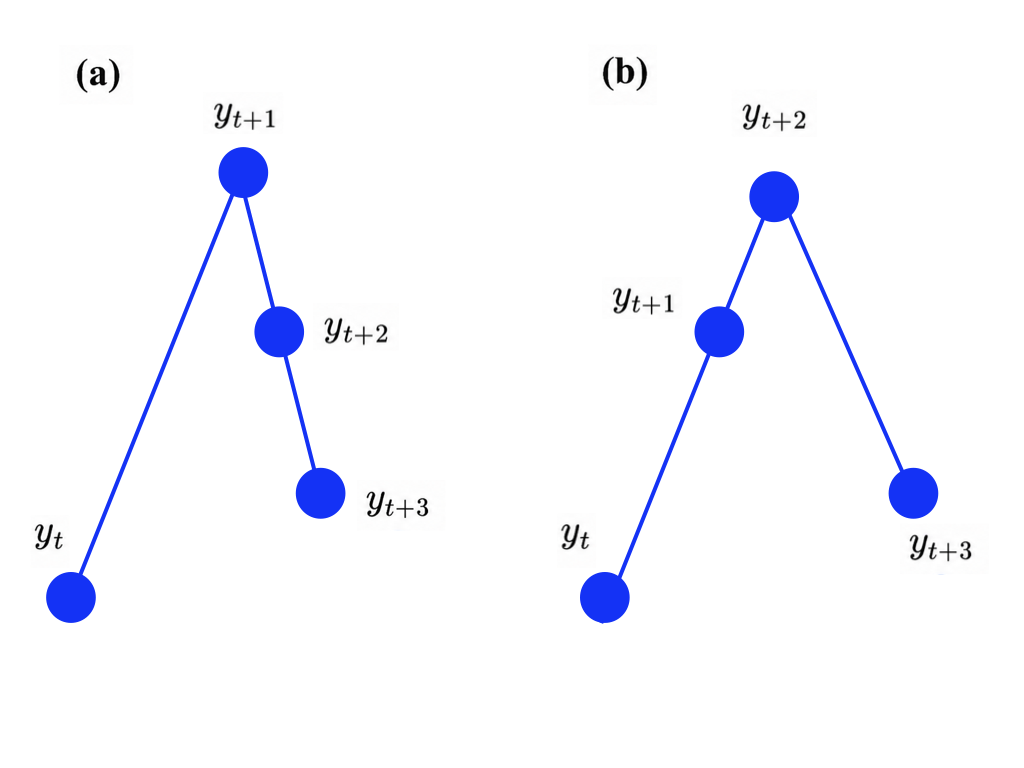}
    \caption{Graphical representation of the ordinal patterns (a) $(0321)$ and (b) $(0231)$}
    \label{fig:graph_rep}
\end{figure}

On the other hand, we observe a subset of ordinal patterns whose probabilities vary systematically with the parameter $q$. Interestingly, these are precisely the same patterns for which the associated integrals cannot be resolved through symmetry arguments alone and instead depend explicitly on the functional form of the underlying distribution. To further investigate this dependence, Fig.~\ref{fig:sim_teo} compares the probabilities of the ordinal patterns $(0231)$ and $(0321)$, obtained from simulated time series and from the numerical evaluation of their corresponding multidimensional integrals (see Eq.~(\ref{eq_int_0231}) for the pattern $(0231)$). An excellent agreement between both approaches is observed across the full range of explored $q$ values, thereby providing strong validation of both the analytical formulation and the numerical implementation. 
The different behavior of the patterns $(0321)$ and $(0231)$ can be directly understood from their associated inequality constraints. For the integrated process, the pattern $(0321)$ corresponds to
\begin{equation}
0 < x_1+x_2+x_3 < x_1+x_2 < x_1,
\end{equation}
which implies
\begin{equation}
x_2<0,\qquad x_3<0,\qquad x_1>-x_2-x_3,
\end{equation}
that corresponds to a large positive increment followed by two consecutive negative steps. This relation is more clearly illustrated in the graphical representation shown in Fig.~\ref{fig:graph_rep}. As $q$ increases, the $q$-Gaussian distribution develops heavier tails, increasing the probability of extreme increments. Consequently, trajectories dominated by a single large fluctuation become more likely.

In contrast, the pattern $(0231)$ satisfies the inequalities given by Eq.~(\ref{inq_0231}), which corresponds to a trajectory with two consecutive positive steps, followed by a negative fluctuation partially compensated by the previous positive increments. Therefore, this pattern depends on a relatively balanced combination of increments. Such trajectories are comparatively less favored in the heavy-tailed regime.

For $q>5/3$, the variance of the $q$-Gaussian distribution diverges and is therefore not defined. As a consequence, the distribution develops extremely heavy tails, leading to a rapidly increasing probability of very large increments. In this regime, the integrated trajectories become dominated by rare extreme events, causing a strong concentration of probability on a reduced subset of ordinal patterns. Consequently, the convergence of the simulated ordinal probabilities becomes significantly slower, since a finite number of realizations is no longer sufficient to properly sample these rare fluctuations. For this reason, in the present work we restrict our analysis to the range $q\leq 2.5$, where reliable statistical convergence is achieved.

\subsection{Asymmetric Distributions}

For symmetric distributions centered around zero, the construction of the integrated process (see Eq.~(\ref{walk})) is essentially independent of the choice of the centering parameter, since $\langle x_t \rangle=0$. As a consequence, the increments preserve the symmetry of the original distribution and do not introduce systematic distortions in the ordinal probabilities. However, this situation changes when asymmetric distributions are considered. In heavy-tailed asymmetric regimes, the mean can be strongly shifted by rare extreme events, causing the integrated process to acquire an artificial drift when mean-centering is applied. As will be shown below, this effect may produce an overestimation of specific ordinal patterns and generate apparent ordinal asymmetries that do not originate from the intrinsic structure of the underlying distribution.

\begin{figure*}[t!]
    \centering
    \includegraphics[width=\textwidth]{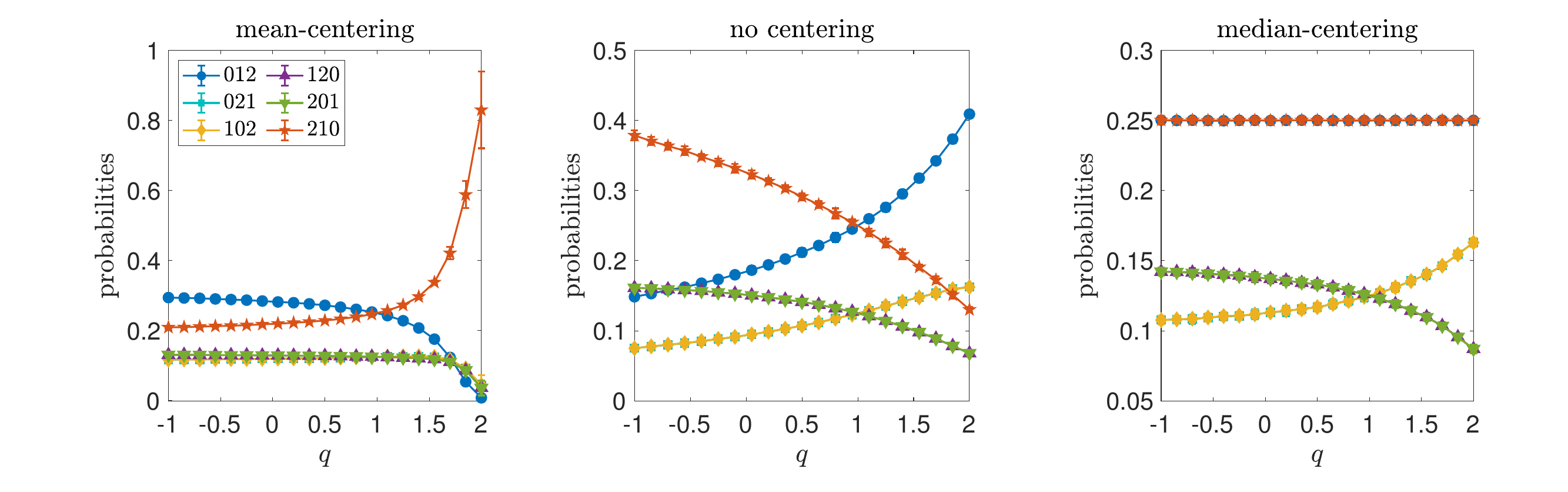}
    \caption{Ordinal pattern probabilities ($D=3$) obtained from the numerical simulations of random walks with increments drawn from a asymmetric distribution using mean-centering (left panel), no centering (middle panel) and median-centering (right panel). Mean and standard deviation (as errorbars) from 20 realizations are plotted. Errorbars are not visible because the standard deviation is smaller than the symbol size.}
    \label{fig:com_sim_asy_q}
\end{figure*}

This effect can be explicitly observed in the analytical inequalities defining the ordinal patterns. In the general case, the integrated process can be written as
\begin{equation}
y_t=\sum_{i=1}^{t}(x_i-\gamma),
\end{equation}
where \(\gamma\) represents the selected centering parameter. For instance, for the ascending pattern $(012)$ in the case $D=3$, the ordering condition is
\begin{equation}
y_t<y_{t+1}<y_{t+2}.
\end{equation}
When a centering parameter \(\gamma\) is introduced, one has
\begin{equation}
\begin{aligned}
y_{t}   &= 0, \\
y_{t+1} &= x_1-\gamma, \\
y_{t+2} &= x_1+x_2-2\gamma.
\end{aligned}
\end{equation}
The ordering condition then becomes
\begin{equation}\label{des_centering}
0<x_1-\gamma<x_1+x_2-2\gamma,
\end{equation}
which yields
\begin{equation}
x_1>\gamma,
\qquad
x_2>\gamma.
\end{equation}
This shows that the choice of the centering parameter directly modifies the inequalities defining the ordinal patterns and, consequently, their associated integration domains. In asymmetric heavy-tailed distributions, the mean is shifted toward the heavy tail due to the contribution of rare extreme events. As a result, mean-centering introduces an artificial drift in the integrated process, leading to an overestimation or underestimation of the ascending and descending ordinal patterns depending on the direction of the asymmetry. Therefore, the observed ordinal probabilities may reflect the bias introduced by the centering procedure rather than the genuine structure of the underlying distribution.

To illustrate this effect, we perform a numerical analysis using time series drawn from an asymmetric distribution obtained by joining two half-distributions at the origin. The left branch ($x < 0$) is drawn from a standard Gaussian with unnormalized density $g_L(x) = e^{-x^2/2}$, while the right branch ($x > 0$) follows a $q$-Gaussian with unnormalized density $g_R(x) = \left[1 - (1-q)\frac{x^2}{2}\right]_+^{\frac{1}{1-q}}$. Crucially, both components evaluate to unity at the origin, $g_L(0) = g_R(0) = 1$, ensuring that the resulting density is continuous at the junction point. The joint distribution is then defined as
\begin{equation}
    f(x) =
    \begin{cases}
        \dfrac{g_L(x)}{Z_L + Z_R}, & x \leq 0, \\[10pt]
        \dfrac{g_R(x)}{Z_L + Z_R}, & x > 0,
    \end{cases}
\end{equation}
where $Z_L = \int_0^{\infty} e^{-t^2/2}\, dt =\sqrt{\pi/2}$ and $Z_R = \int_0^{\infty} g_R(t)\, dt$ are the normalization integrals of each half-branch. The probability that a sample falls on the negative (positive) side is $P_L = Z_L/(Z_L + Z_R)$ ($P_R = Z_R/(Z_L + Z_R)$), determined entirely by the relative mass of each branch. For $q = 1$, both branches reduce to halves of a standard Gaussian, so $Z_L = Z_R$ and the distribution is fully symmetric. As $q$ increases above 1, the right tail becomes heavier ($Z_R > Z_L$) and positive fluctuations grow more probable, whereas for $q < 1$ the right branch acquires compact support ($Z_R < Z_L$), making the Gaussian branch dominant. 

Figure~\ref{fig:com_sim_asy_q} shows the comparison of three centering parameters (left panel) $\gamma=\langle x_t \rangle$,  (middle panel) $\gamma=0$, that implies using the raw data with no centering, and (right panel) $\gamma=\mathrm{median}(x_t)$. For $q=1$, the distribution reduces to the Gaussian case, recovering the well-known ordinal pattern probabilities (see Eqs.~(\ref{probD3A}) and (\ref{probD3B})). Since the distribution becomes symmetric around zero, the mean, median, and zero-centering are equivalent, leading to identical ordinal probabilities independently of the centering procedure used. 

For $q>1$, and when the traditional selection of the mean-centering is used, extreme positive events strongly shift the mean toward larger values, causing most transformed increments to become effectively negative, as evidenced in Eq.~(\ref{des_centering}). In other words, most observations become relatively small after centering, artificially favoring the descending ordinal pattern, as shown in the left panel in Fig.~\ref{fig:com_sim_asy_q}. Note that a heavy left tail would produce the opposite effect. Interestingly, a similar effect can also arise for symmetric $q$-Gaussian distributions when $q>2$. In this regime, the tails become sufficiently heavy that the mean diverges. Consequently, rare but extremely large fluctuations (outliers) can dominate finite samples, effectively shifting the sample mean and introducing an artificial bias in the ordinal pattern probabilities, despite the underlying distribution being perfectly symmetric. On the other hand, when no centering is applied, the process preserves the intrinsic asymmetry of the underlying distribution. Large positive fluctuations naturally favor increasing trajectories, leading to an overestimation of the ascending pattern, as observed in the middle panel in Fig.~\ref{fig:com_sim_asy_q}.

For $q<1$, the $q$-Gaussian distribution develops compact support and extreme fluctuations progressively disappear. At the same time, the asymmetry of the constructed distribution becomes effectively dominated by the Gaussian branch on the negative side, which now extends over a broader range than the bounded positive $q$-Gaussian component. As a consequence, the mean becomes less affected by rare events and recovers statistical stability. In this regime, the artificial drift introduced by mean-centering is significantly reduced, leading to an inversion of the effect observed for $q>1$. Similarly, when no centering is applied, the absence of large positive excursions prevents the walk from developing the strong upward bias observed in the heavy-tailed regime, reducing the overestimation of the ascending ordinal pattern.

In contrast, median-centering provides a considerably more stable representation of the ordinal structure in asymmetric distributions. Since the median depends only on the ordering of the observations and not on their magnitude, it does not introduce the artificial drift observed with the mean, that depends linearly on the magnitude of all observations~\cite{Huber2009}. As shown in the right panel in Fig.~\ref{fig:com_sim_asy_q}, the probabilities of the ascending and descending patterns remain constant with $q$ and equal to $1/4$, consistently reflecting the absence of temporal correlations in the data. Instead, the effect of the asymmetry is transferred to the "turning" ordinal patterns, which are associated with local convex or concave structures in the trajectory. These patterns become sensitive to the imbalance between positive and negative fluctuations, allowing the ordinal probabilities to characterize the intrinsic asymmetry of the distribution without contamination from artificial trends. The present conclusions can be naturally extended to other values of the embedding dimension $D$, although these results are not shown here for visualization clarity of the figures.

These results are consistent with the proposal of Bandt and Shiha, who showed that specific linear combinations of ordinal pattern probabilities for $D=3$ can be interpreted as measures of persistence and symmetry in time series~\cite{bandt2007order,bandt2023statistics}. In particular, monotonic patterns are associated with temporal persistence, whereas turning patterns quantify deviations from spatially symmetric processes. 

\section{Empirical Application}

In this section, we extend the analysis to empirical data in order to evaluate whether the ordinal structures observed in real systems are consistent with the theoretical distributions previously obtained. To perform a more global comparison between ordinal distributions, instead of analyzing individual pattern probabilities separately, we employ the Jensen-Shannon distance as a quantitative measure of dissimilarity between two ordinal probability distributions~\cite{zunino2022permutation}. This approach allows us to characterize the overall agreement between empirical and theoretical results in a more robust way.

\subsection*{Financial markets time series}

Financial markets are one of the paradigmatic examples of complex systems, where the collective interaction of many agents generates highly fluctuating dynamics. A central pillar of financial economics is the efficient-market hypothesis (EMH), which states that asset prices fully and instantaneously incorporate all available information. Under this assumption, price dynamics are often modeled as a random walk with independent and identically distributed increments and no predictable structure. However, while the EMH provides a framework for understanding the temporal unpredictability of financial returns, it does not imply that their fluctuations must follow Gaussian statistics. On the contrary, a large body of empirical evidence has shown that financial time series strongly deviate from Gaussian behavior, exhibiting heavy-tailed return distributions and an enhanced probability of extreme events~\cite{mandelbrot1963new,gopikrishnan1998inverse,nayak2021computational,gabaix2003theory,drozdz2007stock}. In particular, the seminal work of Gopikrishnan et al.~\cite{gopikrishnan1998inverse} revealed the existence of the so-called inverse cubic law for financial returns, subsequently motivating several theoretical efforts aimed at explaining its origin and apparent universality~\cite{gabaix2003theory}. Despite the extensive evidence of non-Gaussian behavior in financial returns, there is still no consensus on which probability distribution provides the most appropriate description of their statistical properties~\cite{de2023modeling}. More recently, Nayak et al.~\cite{nayak2021computational} employed the framework of non-extensive Tsallis statistics to model the non-Gaussian behavior of financial returns, showing that high-frequency returns are well described by $q$-Gaussian distributions with $q \approx 5/3$, thereby reproducing both the observed heavy tails and the inverse cubic law. Similar power-law tails have subsequently been reported in a variety of stock markets worldwide, suggesting that heavy-tailed fluctuations constitute a universal feature of financial dynamics across different economic environments~\cite{drozdz2007stock}.

\begin{figure}[t!]
    \centering
    \includegraphics[width=\linewidth]{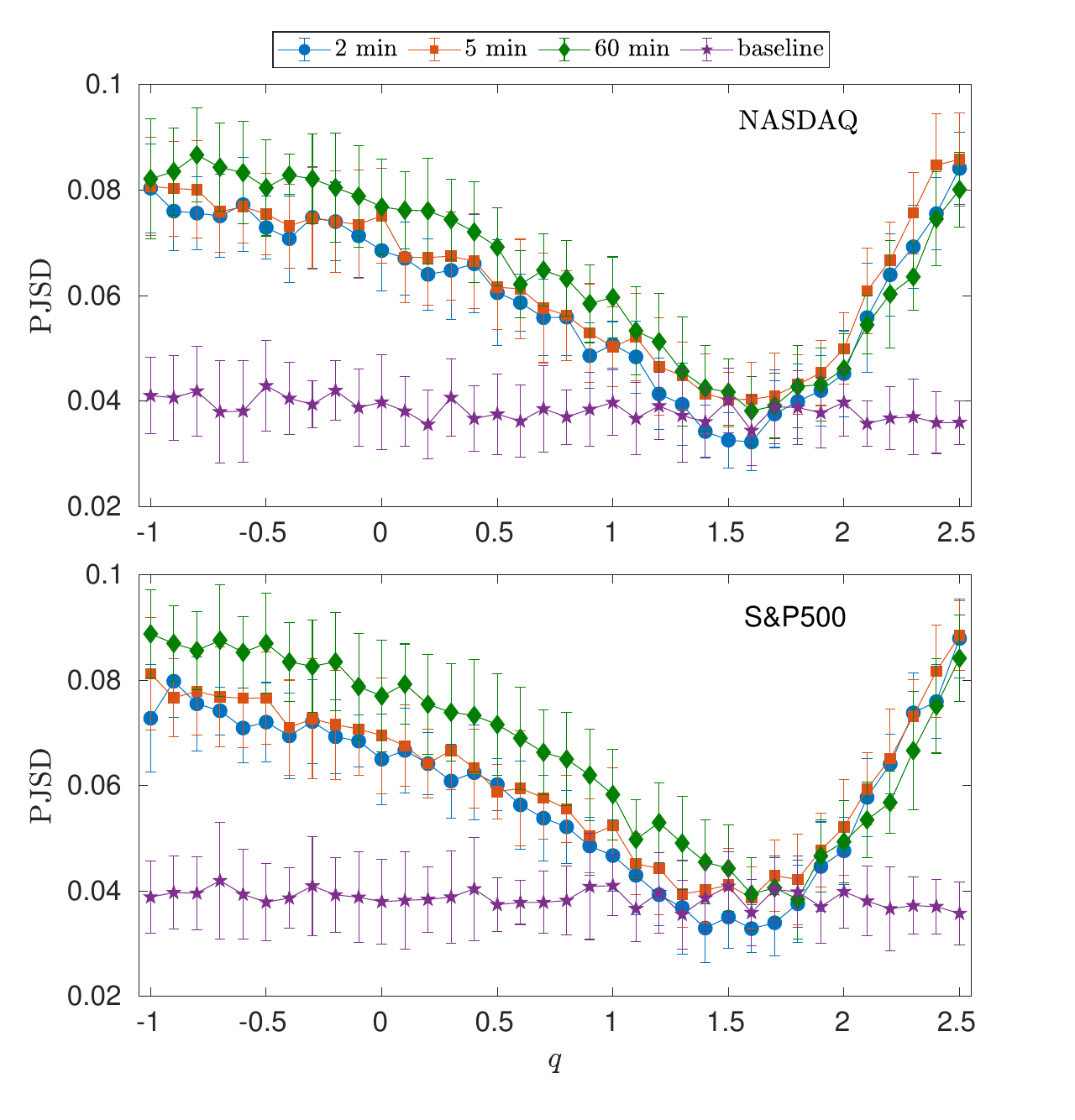}
    \caption{Permutation Jensen-Shannon Distance (PJSD) between the empirical ordinal distributions obtained from the NASDAQ (top panel) and S$\&$P500 (bottom panel) walks and the ones obtained from simulated walks generated with $q$-Gaussian increments for $q \in [-1,2.5]$ with a step of 0.1. Results correspond to ordinal patterns of dimension $D=4$. Symbols correspond to different sampling frequencies: 2 min (blue circles), 5 min (red squares), and 60 min (green triangles). The purple pentagrams indicate the PJSD between two independent realizations of the same $q$-Gaussian random walk, providing the effective zero-distance reference associated with finite-sample effects.
}
    \label{fig:SP500_NASDAQ}
\end{figure}

In financial time series analysis, a standard approach is to study the so-called logarithmic returns instead of raw prices, since returns remove long-term trends and provide a more stationary representation of the underlying dynamics. Given a price series $p_t$, the logarithmic return is defined as
\begin{equation} \label{eq:logreturns}
r_t=\log\left(\frac{p_{t+1}}{p_t}\right).
\end{equation}

We analyzed high-frequency closing-price time series of the S$\&$P500 and NASDAQ index obtained from the Yahoo Finance database. Three sampling frequencies were considered: 2-minute, 5-minute, and 60-minute intervals. The corresponding datasets cover the periods from 21 May 2025 to 10 July 2025 (2-minute sampling), from 14 April 2025 to 10 July 2025 (5-minute sampling), and from 11 August 2022 to 10 July 2025 (60-minute sampling). These sampling frequencies were selected as a compromise between temporal resolution and data availability, ensuring sufficiently long time series to reliably estimate the probabilities for $D=4$. Logarithmic returns were computed from the closing-price series according to Eq.~(\ref{eq:logreturns}). To isolate the effects of the underlying probability distribution, the return series were randomly shuffled prior to the analysis, thereby removing any temporal dependencies while preserving the empirical distribution of fluctuations. Finally, median-centered random walks were then constructed from the shuffled returns.

Figure~\ref{fig:SP500_NASDAQ} shows the Permutation Jensen-Shannon Distance (PJSD) for $D=4$, between the empirical ordinal distributions and the ones obtained from simulated walks with increments from $q$-Gaussian distribution, as a function of $q$. The top and bottom panels correspond to the NASDAQ and S$\&$P500 indices, respectively. A well-defined minimum is observed around $q \in [1.4,1.7]$ for both financial indices and across all sampling frequencies. This minimum reaches the statistical baseline, corresponding to the effective zero distance expected from finite-size effects~\cite{zunino2022permutation,olivares2023markov}. Although an independent baseline was computed for each sampling frequency, only one is displayed because they are statistically indistinguishable, even though different sampling frequencies lead to different time-series lengths due to the fixed observation period. Interestingly, the optimal $q$ interval
contains the value $q=5/3$, which has been repeatedly associated with the non-Gaussian statistics of financial returns and the inverse cubic law reported in empirical market data~\cite{gopikrishnan1999scaling,drozdz2007stock,nayak2021computational}.

\section{Conclusions}

In this work, we have shown that the ordinal structure of integrated stochastic processes provides direct information about the amplitude distribution of the underlying fluctuations, overcoming one of the classical limitations of ordinal pattern analysis. While ordinal probabilities computed from i.i.d. observations are insensitive to the underlying probability density function, the integration of the fluctuations transforms amplitude statistics into geometric constraints on the resulting trajectories, making distributional properties accessible through ordinal analysis. Analytical and numerical results show that, for $D=4$, some ordinal pattern probabilities are fully determined by symmetry, whereas the remaining ones encode the underlying increment distribution.

We have further demonstrated that the construction of the integrated process is crucial for accurately characterizing distribution of the fluctuations. In particular, for asymmetric heavy-tailed distributions, conventional mean-centering introduces artificial trends that substantially distort the ordinal probabilities. Median-centering, by contrast, provides a robust alternative that preserves the intrinsic statistical properties of the fluctuations while avoiding biases caused by extreme events. These results highlight that an appropriate centering strategy is essential when applying ordinal analysis to integrated processes with heavy-tailed distributions.

Finally, we have validated the proposed framework using high-frequency financial time series. The ordinal distributions of integrated logarithmic returns were found to be in closest agreement with simulated $q$-Gaussian walks for values of $q$ around $5/3$, consistently recovering the heavy-tailed behavior associated with the inverse cubic law reported in financial markets. Beyond finance, the proposed framework is completely general and can be applied to any stochastic process in which the statistical properties of the increments are of interest. By establishing a direct connection between amplitude distributions and ordinal statistics through integration, this work broadens the scope of ordinal methods and provides a simple methodology for characterizing non-Gaussian stochastic dynamics.

\appendix
\section{Explicit integral expressions for $D=4$ ordinal pattern probabilities}\label{Apendice}

The complete ordinal pattern distribution for $D=4$ consists of two subsets. The first comprises the probabilities that are completely determined by symmetry and are given explicitly in Eqs.~(\ref{patrones1}). These results follow solely from the geometry of the integration domains, whose limits involve only adjacent increments and therefore do not require combinations of multiple variables. Consequently, the corresponding probabilities are independent of the particular form of the increment distribution and can be obtained exactly from symmetry arguments. For reference, the explicit integral representations of six ordinal patterns are given below
\begin{align*}
P(0123) &=
\int_{0}^{\infty}\!\!\int_{0}^{\infty}\!\!\int_{0}^{\infty}
f(x_1)f(x_2)f(x_3)\,dx_3\,dx_2\,dx_1
=\frac18,\\[2mm]
P(0132) &=
\int_{0}^{\infty}\!\!\int_{0}^{\infty}\!\!\int_{-x_2}^{0}
f(x_1)f(x_2)f(x_3)\,dx_3\,dx_2\,dx_1
=\frac1{16},\\[2mm]
P(0213) &=
\int_{0}^{\infty}\!\!\int_{-x_1}^{0}\!\!\int_{-x_2}^{\infty}
f(x_1)f(x_2)f(x_3)\,dx_3\,dx_2\,dx_1
=\frac1{24},\\[2mm]
P(0312) &=
\int_{0}^{\infty}\!\!\int_{-x_1}^{0}\!\!\int_{0}^{-x_2}
f(x_1)f(x_2)f(x_3)\,dx_3\,dx_2\,dx_1
=\frac1{48},\\[2mm]
P(1023) &=
\int_{-\infty}^{0}\!\!\int_{-x_1}^{\infty}\!\!\int_{0}^{\infty}
f(x_1)f(x_2)f(x_3)\,dx_3\,dx_2\,dx_1
=\frac1{16},\\[2mm]
P(1203) &=
\int_{0}^{\infty}\!\!\int_{-\infty}^{-x_1}\!\!\int_{-x_2}^{\infty}
f(x_1)f(x_2)f(x_3)\,dx_3\,dx_2\,dx_1
=\frac1{48}.
\end{align*}
The remaining six probabilities follow directly from the time-reversal symmetry, $P(\pi)=P(\pi^{\mathrm{rev}})$, where $\pi^{\mathrm{rev}}$ denotes the permutation obtained by reversing the temporal order of the pattern. Consequently,
\begin{align*}
P(3210)&=P(0123),\\
P(2310)&=P(0132),\\
P(3120)&=P(0213),\\
P(2130)&=P(0312),\\
P(3201)&=P(1023),\\
P(3021)&=P(1203).
\end{align*}

The second subset contains the twelve probabilities whose integration domains involve combinations of consecutive increments, such as $x_1+x_2$. Consequently, these probabilities are not fixed by symmetry and depend explicitly on the functional form of the increment distribution $f(x)$, and they must therefore be evaluated numerically. Considering again that the time-reversal symmetry of random walks with independent increments drawn from a symmetric distribution, every ordinal pattern has the same probability as its reversed counterpart. As a result, only six independent probabilities need to be evaluated explicitly. These independent probabilities are given by the following triple integrals:
\begin{align*}
P(0231) &=
\int_{0}^{\infty}\!\!\int_{0}^{\infty}\!\!\int_{-(x_1+x_2)}^{-x_2}
f(x_1)f(x_2)f(x_3)\,dx_3\,dx_2\,dx_1,
\\[2mm]
P(0321) &=
\int_{0}^{\infty}\!\!\int_{-x_1}^{0}\!\!\int_{-(x_1+x_2)}^{0}
f(x_1)f(x_2)f(x_3)\,dx_3\,dx_2\,dx_1,
\\[2mm]
P(1032) &=
\int_{-\infty}^{0}\!\!\int_{-x_1}^{\infty}\!\!\int_{-(x_1+x_2)}^{0}
f(x_1)f(x_2)f(x_3)\,dx_3\,dx_2\,dx_1,
\\[2mm]
P(1302) &=
\int_{0}^{\infty}\!\!\int_{-\infty}^{-x_1}\!\!\int_{-(x_1+x_2)}^{-x_2}
f(x_1)f(x_2)f(x_3)\,dx_3\,dx_2\,dx_1,
\\[2mm]
P(2013) &=
\int_{-\infty}^{0}\!\!\int_{0}^{-x_1}\!\!\int_{-(x_1+x_2)}^{\infty}
f(x_1)f(x_2)f(x_3)\,dx_3\,dx_2\,dx_1,
\\[2mm]
P(2103) &=
\int_{-\infty}^{0}\!\!\int_{-\infty}^{0}\!\!\int_{-(x_1+x_2)}^{\infty}
f(x_1)f(x_2)f(x_3)\,dx_3\,dx_2\,dx_1,
\end{align*}
and the remaining six probabilities are determined by time-reversal symmetry as follows:
\begin{align*}
P(1320)&=P(0231),\\
P(1230)&=P(0321),\\
P(2301)&=P(1032), \\
P(2031)&=P(1302),\\
P(3102)&=P(2013), \\
P(3012)&=P(2103).
\end{align*}

\begin{acknowledgments}
AMR acknowledges support from the SURF@IFISC (Summer Undergraduate Research Fellowships at IFISC) program. LZ gratefully acknowledges financial support from Consejo Nacional de Investigaciones Cient\'ificas y T\'ecnicas (CONICET), Argentina. FO acknowledges the Spanish State Research Agency through the Mar\'ia de Maeztu project CEX2021-001164-M funded by the  MCIN/AEI/10.13039/501100011033. 
\end{acknowledgments}

\bibliography{biblio}

\end{document}